\documentclass[preprint,3p,12pt]{elsarticle}

\usepackage{amssymb}
\usepackage{pdfsync}
\usepackage{bm}
\usepackage{subfigure} 
\usepackage{amssymb}
\usepackage{psfrag}
\usepackage{amsmath}
\usepackage{verbatim}
\usepackage{my_symbols}
\usepackage{amsfonts}
\usepackage{graphicx}
\usepackage{tikz}
\usepackage{upgreek}
\usepackage{textcomp}
\usetikzlibrary{calc,3d}
\usetikzlibrary{arrows}

\journal{Solid State Physics}

\begin{document}

\begin{frontmatter}



\title{Spin Wave Excitation in Magnetic Insulator Thin Films by Spin-Transfer Torque}


\author[a,b]{Jiang Xiao}
\author[c]{Yan Zhou} 
\author[d,e]{Gerrit E. W. Bauer}

\address[a]{Department of Physics and State Key Laboratory of Surface Physics, Fudan University, Shanghai, China}

\address[b]{Center for Spintronic Devices and Applications, Fudan University, Shanghai, China}

\address[c]{Department of Physics, The University of Hong Kong, Hong Kong, China}

\address[d]{Institute for Materials Research $\&$ WPI-AIMR, Tohoku University, Sendai 980-8557, Japan}

\address[e]{Kavli Institute of NanoScience, Delft University of Technology, 2628 CJ Delft, The Netherlands}

\begin{abstract}

    We describe excitation of dipole-exchange spin waves in insulating magnetic thin films by spin current injection at the surface of the film. An easy-axis magnetic surface anisotropy can induce a non-chiral surface spin wave mode with penetration depth inversely proportional to the strength of the surface anisotropy, which strongly reduces the critical current and enhances the excitation power. The importance of the interface spin wave modes on the excitation spectrum is reduced by spin pumping, which depends on the quality of the interface as expressed by the spin mixing conductance.
\end{abstract}


\end{frontmatter}

\tableofcontents

% \linenumbers




%###############################################################

%=========================================================================
\section{Introduction \& background} 

Since its prediction \cite{berger_emission_1996, slonczewski_current-driven_1996}, the spin-transfer torque has been studied extensively both experimentally and theoretically in nano-structures, including magnetic point contacts, spin valves, magnetic tunnel junctions, magnetic thin film structures containing magnetization textures such as magnetic domain walls, vortices, skyrmions, $\etc$. \cite{brataas_current-induced_2012} 
%(and references therein).  
%\cite{tsoi_excitation_1998,myers_current-induced_1999,sun_spin-current_2000,slonczewski_currents_2002,katine_current-driven_2000,ozyilmaz_current-induced_2003,kiselev_microwave_2003,brataas_finite-element_2000,xia_spin_2002,stiles_anatomy_2002,xiao_boltzmann_2004,xiao_macrospin_2005,xiao_spin-transfer_2006,huai_observation_2004,sankey_measurement_2008,kubota_quantitative_2008,theodonis_anomalous_2006,xiao_spin-transfer_2008,oh_bias-voltage_2009,freitas_observation_1985,yamaguchi_real-space_2004,vernier_domain_2004,klaui_controlled_2005,zhang_roles_2004,thiaville_micromagnetic_2005,tatara_theory_2004,heyne_direct_2010}
When the torque is sufficiently large at a certain critical current, the magnetization is set into motion. In nanostructures the current-induced dynamics is often well described by the macrospin model that assumes uniform collective motion of the entire magnetization. On the other hand, by employing the (inverse) spin Hall effect \cite{saitoh_conversion_2006}, a uniform spin current can be injected into a macroscopically large area, thereby possibly exciting the whole spectrum of spin waves.

\begin{figure}[t]
\centering
        \includegraphics[width=\textwidth]{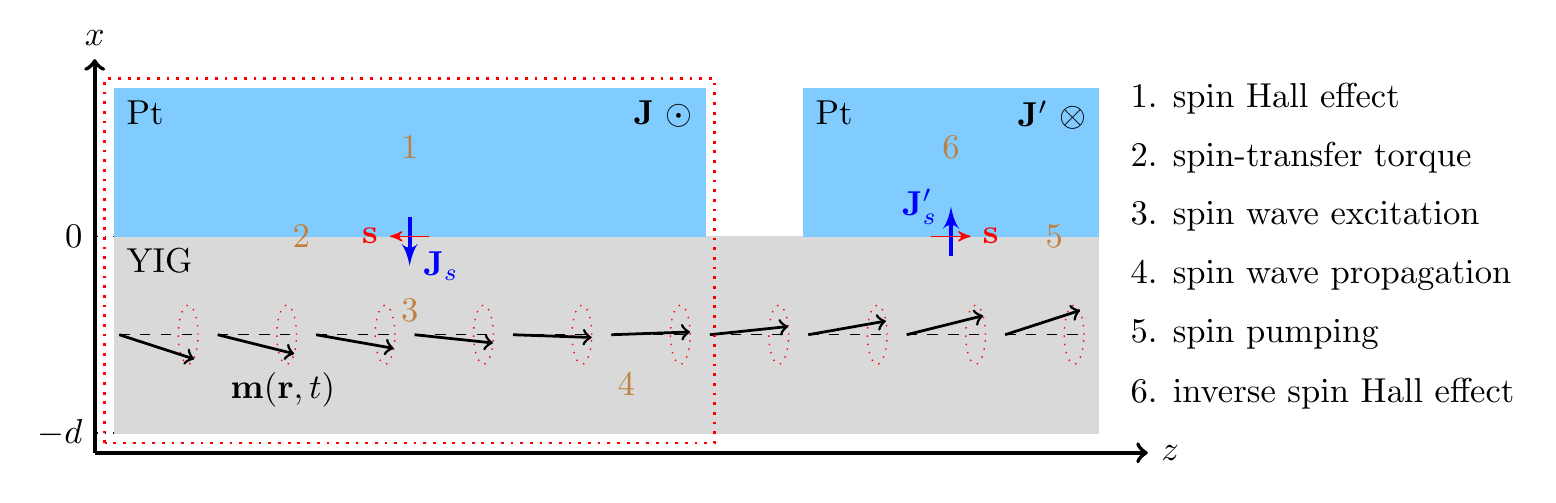} 
\caption{Schematic side view of the experimental setup (Ref. \cite{kajiwara_transmission_2010}). The six consecutive processes: 1) the spin Hall effect generates a spin current in the Pt on the left, 2) the spin current exerts a spin-transfer torque on the magnetization at the interface, 3) spin wave is excited in YIG when the spin-transfer torque overcomes the damping, 4) the excited spin wave propagates within the film, 5) the spin wave pumps a spin current into the Pt contact on the right, 6) the inverse spin Hall effect converts the pumped spin current into a charge current. }
\label{fig:setup}
\end{figure}

A spintronic circuit consists of devices that perform operations and interconnects that transport information. Spin information communication over long distances requires weak coupling with the outside world, while spintronic devices require strong coupling with an external control parameter in order to efficiently manipulate the spins. Materials and designs that fulfil these conflicting demands are highly desirable. Spin waves can propagate over much longer distances than diffusive spin currents,  in materials such as the magnetic insulator Yttrium-Iron-Garnet (YIG) \cite{serga_yig_2010}. Therefore it appears attractive to transmit spin information via spin waves rather than particle-based spin currents \cite{Khitun2007}. To interface a spintronic circuit, an electrical signal has to be encoded into a spin signal, and spin information has to be read out as a voltage signal. This scenario has been demonstrated by Kajiwara \etal \cite{kajiwara_transmission_2010} using a Pt-YIG-Pt structure as schematically shown in \Figure{fig:setup}, in which the encoding is realized by exciting spin waves in a magnetic thin film electrically using the spin Hall effect and spin-transfer torque \cite{berger_emission_1996, slonczewski_current-driven_1996}. The readout is realized by the inverse effect ---  spin pumping \cite{tserkovnyak_enhanced_2002}, which converts spin waves back into a spin current and via the inverse spin Hall effect, a voltage \cite{kajiwara_transmission_2010, madami_direct_2011, wang_control_2011}. The full scenario can be described by six consecutive processes: i) the spin Hall effect in N generates a spin current $\JJ_s$ polarized long $-\hzz$ by a charge current $J$ flowing in $-\hyy$-direction that is absorbed by the YIG film, ii) the spin current exerts a torque on the magnetization in YIG at the interface, iii) spin wave is excited when the spin-transfer torque overcomes the damping, iv) the excited spin waves propagate to the Pt detector on the right, v) the spin wave reaching the second interface contact pumps a spin current into the right Pt, vi) the pumped spin current is converted back into a charge current via the inverse spin Hall effect. 

 In the experiment, spin waves were excited in a $d = 1.3~\upmu$m-thick YIG film by a threshold charge current of $J_c {\sim} 10^9$ A/m$^2$ in a 10 nm thick Pt overlayer \cite{kajiwara_transmission_2010}. This value is much less than expected for a volume spin wave excitation that in the macrospin model  is estimated to be $J_c = (1/{\theta}_H) e{\alpha}{\omega}M_sd/ {\gamma}{\hbar}~{\sim}~ 5{\times}10^{11}$ A/m$^2$, where $e$ and ${\gamma}=e/2m$ are the electron charge and gyromagnetic ratio, respectively, and we used the material parameters in Table \ref{tab:param} for the ferromagnetic resonance frequency ${\omega}$, the spin Hall angle  ${\theta}_H$, magnetic damping ${\alpha}$, and saturation magnetization $M_s$.  

In this Chapter, we  focus on the excitation part, \ie the processes inside the dashed box in \Figure{fig:setup}. We describe the spin wave dispersion and excitation in insulating magnetic thin films by the spin-transfer torque in the linear response regime, obtaining the threshold currents for excitation of various spin wave modes and the low-excitation power spectrum. 

%------------------
%------------------

%=========================================================================
\section{Spin current induced magnetization dynamics} 

We study the spin wave excitation in a bilayer structure as shown in the dotted box region in \Figure{fig:setup}, in which a non-magnetic (N) metallic thin film is in contact with a ferromagnetic insulator (FI) with equilibrium magnetization along the $\hzz$-direction. The charge current in Pt flows normal the displayed cross section.

The spatial dependent dynamics of the magnetization unit vector $\mm(\rr,t)$ within the ferromagnetic film is described by the Landau-Lifshitz-Gilbert-Slonczewski (LLGS) equation: 
%-----------------------------------------
\begin{equation}
    \dmm = -{\gamma}~\mm{\times}\midb{\HH_0 + \HH_s + {A_{\rm ex}\ov {\gamma}}{\nabla}^2\mm + \hh} + {\alpha}~\mm{\times}\dmm  + 
    {{\gamma}\ov M_s}\smlb{\btau_{\rm STT}+\btau_{\rm SP}},
\label{eqn:llgs}
\end{equation}
%----------------------------------------- 
where $\HH_0$ includes the external and internal magnetic field, $\HH_s$ is the surface anisotropy field, $A_{\rm ex}$ is the exchange constant, $\hh(\rr,t)$ is the dipolar magnetic field due to $\mm(\rr,t)$, ${\alpha}$ is the Gilbert damping constant, $\btau_{\rm STT}$ is the spin-transfer torque due to the spin current injection from the top surface and $\btau_{\rm SP}$ is the spin pumping torque due to the magnetization dynamics at the interface.  

Due to the broken symmetry, the spins at the surface experience a modified magnetic anisotropy that is usually uniaxial and can be described by a surface anisotropy field \cite{gurevich_magnetization_1996}:
%----------------------------------------- 
\begin{equation}
    \HH_s = {2K_1\ov M_s}(\mm{\cdot}\nb)\nb,
\label{eqn:Hs}
\end{equation}
%----------------------------------------- 
where $\nb$ is the outward normal as seen from the ferromagnet which can be the easy or hard axis, depending on the sign of the anisotropy constant $K_1$.

At the top surface ($x = 0$, see \Figure{fig:setup}), a spin Hall effect induced spin current density $\JJ_s = J_s\hzz$ polarized along ${\pm}\hzz$ flowing in $-\hxx$ direction gives rise to a spin accumulation in N, which cancels out the longitudinal component of the spin current $\JJ_s$ parallel to $\mm$. The transverse component is absorbed at the surface as spin-transfer torque, which drives the magnetization dynamics \cite{stiles_anatomy_2002}.  The  spin pumping exerts a torque $\btau_{\rm SP}$ on $\mm$. The volume density of the spin transfer torque and spin pumping torque can be written as
%----------------------------------------- 
\begin{equation}
    \btau_{\rm STT} = J_s \mm {\times} \hzz {\times} \mm~{\delta}(x) \qand
    \btau_{\rm SP} = {\hbar}{g_r\ov 4{\pi}} \hzz {\times} \dmm~{\delta}(x)
\label{eqn:STT}
\end{equation}
%----------------------------------------- 
where $g_r$ is the real part of the mixing conductance per area for the Pt/YIG interface and the ${\delta}$-function indicates that both torques are localized at the surface. There is no current-induced torque inside the bulk because the magnet is assumed to be electrically insulating.

%The spin accumulation at $x = 0$ induced by a uniform spin current polarized along the $\hzz$-direction $\JJ_s = J_s\hzz$ flowing in $-\hxx$ direction is approximately ${\delta}{\bm {\mu}} {\simeq} (2e^2{\rho}_N{\lambda}_N/{\hbar})J_s\hzz {\simeq} (3\sqrt{3}{\pi}^2/k_F^2\bar{{\eta}}_{\rm so})J_s\hzz$, where ${\rho}_N, {\lambda}_n, k_F, \bar{{\eta}}_{\rm so}$ are the electrical resistivity, spin-diffusion length, Fermi wave-vector, and the dimensionless spin-orbit coupling parameter of N, respectively, as defined in Ref.  \cite{takahashi_spin_2008}. The spin-transfer torque at the N/FI interface due to ${\delta}{\bm{\mu}}$ is ${\bm {\tau}} = (g_r/4{\pi}A)\mm{\times}{\delta}{\bm {\mu}}{\times}\mm$ with $g_r/A {\sim} k_F^2/4{\pi}$ the mixing conductance per area for perfect interface. Therefore ${\bm{\tau}}{\simeq} {\xi} J_s\mm{\times}\hzz{\times}\mm$ with ${\xi} = 3\sqrt{3}/16\bar{{\eta}}_{\rm so} {\sim} 1$.  Its transverse component is absorbed at the interface thus exerting a torque on the magnetization $\mm$ \cite{stiles_anatomy_2002}. 

Assuming that the bottom surface ($x = -d$) is free from the effects of spin-transfer torque, spin pumping, and surface anisotropy, by integrating \Eq{eqn:llgs} over a volume across the surface $x{\in}[0^-,0^+]$ and $x{\in}[-d^-,-d^+]$, we obtain the boundary conditions for $\mm$ at $x = 0$ and $-d$ \cite{gurevich_magnetization_1996}: 
%-----------------------------------------
\begin{subequations}
\label{eqn:mbc}
\begin{align}
        A_{\rm ex}\mm{\times}{{\partial}\mm\ov{\partial}\nb} - {2{\gamma}K_s\ov M_s}(\mm{\cdot}\nb)\mm{\times}\nb 
        + {{\gamma}J_s\ov M_s}\mm{\times}\hzz{\times}\mm + {{\gamma}{\hbar}g_r\ov  4{\pi}M_s}\hzz{\times}\dmm &= 0 \qat x = 0, \\
        A_{\rm ex}\mm{\times}{{\partial}\mm\ov{\partial}\nb} &= 0 \qat x = -d,
%       \mm{\times}{{\partial}\mm\ov{\partial}\nb} - k_s(\mm{\cdot}\nb)\mm{\times}\nb + k_j\mm{\times}\hzz{\times}\mm = 0, 
\end{align}
\end{subequations}
%-----------------------------------------
with ${\partial}\mm/{\partial}\nb~{\equiv}~(\nb{\cdot}{\nabla})\mm$ and $K_s = {\int}_{0^-}^{0^+} K_1 dx$. We convert surface anisotropy, spin current, and spin pumping parameters into effective wave numbers by defining: 
%-----------------------------------------
\begin{equation}
k_s = {2{\gamma}K_s\ov A_{\rm ex}M_s}, \quad  
k_j = {{\gamma}J_s\ov A_{\rm ex}M_s}, \qand
k_p= {{\gamma}g_r{\hbar}{\omega}_0\ov 4{\pi}A_{ex}M_s}.
\label{eqn:ksjp}
\end{equation}
%----------------------------------------- 
As we assumed complete absorption of the transverse spin current at the interface, $k_j$ literally means the absorbed spin current. Analogously, the wave number  $k_c = {\alpha}({\omega}_0+{\omega}_M/2)d/A_{\rm ex}$ corresponds to the threshold current for exciting the uniform mode of thickness $d$. 

The bulk magnetization inside the film satisfies the LLG equation: 
%-----------------------------------------
\begin{equation}
    \dmm = -{\gamma}~\mm{\times}\midb{\HH_0 + {A_{\rm ex}\ov {\gamma}}{\nabla}^2\mm + \hh} + {\alpha}~\mm{\times}\dmm
    \qfor -d<x<0,
\label{eqn:llg}
\end{equation}
%----------------------------------------- 
where the dipolar magnetic field $\hh(\rr,t)$ satisfies Maxwell's equations in the quasistatic approximation: 
%-----------------------------------------
\begin{subequations}
\label{eqn:h}
\begin{align}
    {\nabla}{\times}\hh(\rr,t) &= 0 , \label{eqn:h1} \\
    {\nabla}{\cdot}\bb(\rr,t) = {\nabla}{\cdot}\midb{\hh(\rr,t)+{\mu}_0M_s\mm(\rr,t)} &= 0 
    \qfor -d \le x \le 0, \label{eqn:h2} \\
    {\nabla}{\cdot}\bb(\rr,t) = {\nabla}{\cdot}\hh(\rr,t) &= 0 \qfor x < -d \qor x > 0, \label{eqn:h3}
\end{align}
\end{subequations}
%----------------------------------------- 
with boundary conditions
%-----------------------------------------
\begin{subequations}
\label{eqn:hbc}
\begin{align}
    \hh_{y,z}(0^-) = \hh_{y,z}(0^+), &\quad 
    \bb_x(0^-) = \bb_x(0^+), \\
    \hh_{y,z}(-d^-) = \hh_{y,z}(-d^+), &\quad
    \bb_x(-d^-) = \bb_x(-d^+). 
\end{align}
\end{subequations}
%----------------------------------------- 
\Eqss{eqn:mbc}{eqn:hbc} completely describe what is called dipolar-exchange spin waves. The method described above extends previous studies by De Wames and Wolfram \cite{de_wames_dipole-exchange_1970} and Hillebrands \cite{hillebrands_spin-wave_1990} by including the current-induced spin-transfer torque, spin pumping, and surface anisotropy. 

In general, the non-linear LLG equation in \Eq{eqn:llg} can not be solved easily. However, focussing on  small deviations of $\mm$ from its equilibrium direction ($\hzz$-direction), \ie $\mm(\rr,t) = m_z\hzz + \mm_{\perp} e^{i{\omega}t}$ with $m_z~{\sim}~1$ and $\mm_{\perp} = m_x\hxx + m_y\hyy$, the corresponding dipolar field: $\hh = h_z\hzz + \hh_{\perp} e^{i{\omega}t}$, and the LLG equation can be linearized in terms of $\mm_{\perp}$ and $\hh_{\perp}$:  
%-----------------------------------------
\begin{equation}
    \mm_{\perp}(\rr) = {\chi}({\omega}) {\gamma}\hh_{\perp}(\rr) \qwith \qquad\\ 
    {\chi}({\omega}) = {1\over \hat{{\Omega}}^2 - {\omega}^2}
    \smatrix{\hat{{\Omega}} & i{\omega} \\ -i{\omega} & \hat{{\Omega}}}
\label{eqn:llg-l}
\end{equation}
%----------------------------------------- 
with $\hat{{\Omega}} = {\omega}_0 + i{\alpha}{\omega} - A_{\rm ex}{\nabla}^2$ and ${\omega}_0 = {\gamma}H_0$. 

\Eq{eqn:h1} implies that the dipolar magnetic field can be expressed in terms of a scalar potential: $\hh(\rr,t) = -{\nabla}{\psi}(\rr,t)$. $\bb = \hh + {\mu}_0M_s\hh$ and $\mm_{\perp} = {\chi} {\gamma}\hh_{\perp}$ can also be written in terms of $\hh$ and thus ${\psi}$. Using scalar potential ${\psi}$, \Eq{eqn:h2} can be written as 
%-----------------------------------------
\begin{subequations}
\label{eqn:psi-doe}
\begin{align}
    \smlb{\hat{{\Omega}}^2-{\omega}^2 + {\omega}_M\hat{{\Omega}}}{\nabla}^2{\psi} - {\omega}_M\hat{{\Omega}}{\partial}_z^2{\psi} &= 0
    \qfor -d \le x \le 0, \label{eqn:psi-in} \\
    {\nabla}^2{\psi} &= 0 \qfor x < -d \qor x > 0. \label{eqn:psi-out}
\end{align}
\end{subequations}
%----------------------------------------- 
Because of the translational symmetry in the lateral direction, we may assume that the scalar potential is a plane wave of the form: 
%-----------------------------------------
\begin{equation}
    {\psi}(x,y,z,t) = 
    \begin{cases}
        {\phi}(0) e^{-qx} e^{-i\qq{\cdot}\sS} e^{i{\omega}t} &\qfor x > 0, \\
        {\phi}(x) e^{-i\qq{\cdot}\sS} e^{i{\omega}t} &\qfor -d \le x \le 0, \\
        {\phi}(-d) e^{q(x+d)} e^{-i\qq{\cdot}\sS} e^{i{\omega}t} &\qfor x < -d.  
    \end{cases}
\label{eqn:psi}
\end{equation}
%----------------------------------------- 
where $\sS = (y, z)$ is the in-plane position and $\qq = (q_y, q_z) = q(\sin{\theta},\cos{\theta})$ with $q = \abs{\qq}$ an in-plane wave vector. Using ${\psi}$ in \Eq{eqn:psi}, \Eq{eqn:psi-out} is valid automatically, and the boundary conditions in \Eq{eqn:mbc} and \Eq{eqn:hbc} become
%-----------------------------------------
\begin{subequations}
\label{eqn:bc-psi}
\begin{align}
        \smlb{1+{{\omega}_M\hat{{\Omega}}\ov \hat{{\Omega}}^2-{\omega}^2}}{\partial}_x{\psi}(x) 
    + {i{\omega}_M{\omega}\ov \hat{{\Omega}}^2-{\omega}^2}{\partial}_y{\psi}(x) {\pm} q{\psi}(x) &= 0 \qfor x = 0, -d, \\
%    \mbox{\Eq{eqn:mbc}}: \quad 
\midb{{\partial}_x +ik_{p}{{\omega}\ov{\omega}_0} - \smatrix{k_s & k_j \\ -k_j & 0} }
%        \smatrix{{\partial}_x - k_s+ik_{p}{{\omega}\ov{\omega}_0} & -k_j \\ k_j & {\partial}_x+ik_{p}{{\omega}\ov{\omega}_0}}
    \midb{ {1\ov \hat{{\Omega}}^2-{\omega}^2}\smatrix{\hat{{\Omega}} & i{\omega} \\ -i{\omega} & \hat{{\Omega}}}
        \smatrix{{\partial}_x \\ {\partial}_y}{\psi}(x)} &= 0 \qfor x = 0, \\
    {\partial}_x \midb{
        {1\ov \hat{{\Omega}}^2-{\omega}^2}\smatrix{\hat{{\Omega}} & i{\omega} \\ -i{\omega} & \hat{{\Omega}}}
        \smatrix{{\partial}_x \\ {\partial}_y}{\psi}(x)} &= 0 \qfor x = -d,
\end{align}
\end{subequations}
%----------------------------------------- 
with ${\omega}_M = {\gamma}{\mu}_0M_s$. With \Eq{eqn:bc-psi}, the whole problem is reduced to finding a solution for ${\psi}$, from which $\mm$ and $\hh$ can be calculated. To this end we have to solve the differential equations in \Eq{eqn:psi-doe} subject to the boundary conditions in \Eq{eqn:bc-psi} \cite{de_wames_dipole-exchange_1970, hillebrands_spin-wave_1990}.

With \Eq{eqn:psi}, \Eq{eqn:psi-in} becomes a 6th-order ordinary differential equation of ${\phi}(x)$: 
%-----------------------------------------
\begin{equation}
    \midb{\smlb{\hat{{\Omega}}^2 - {\omega}^2 + {\omega}_M\hat{{\Omega}}}({\partial}_x^2 - q^2) + {\omega}_M\hat{{\Omega}}q_z^2}{\phi}(x) = 0
    \qwith \hat{{\Omega}} = {\omega}_0 + i{\alpha}{\omega} - A_{\rm ex}({\partial}_x^2 - q^2),
\label{eqn:phi-doe}
\end{equation}
%-----------------------------------------
which has the general solution
%-----------------------------------------
\begin{equation}
        {\phi}(x) = {\sum}_{j=1}^3\midb{a_je^{iq_x^{(j)}x}+b_je^{-iq_x^{(j)}(x+d)}},
%       = {\sum}_{j=1}^3\smatrix{e^{iq_x^{(j)}x} & e^{-iq_x^{(j)}(x+d)}} \smatrix{a_j \\ b_j}
\label{eqn:phix}
\end{equation}
%-----------------------------------------
where $q_x^{(j)}$ is the $j$-th solution of
%-----------------------------------------
\begin{equation}
    \smlb{{{\Omega}}^2 - {\omega}^2 + {\omega}_M{\omega}}(q_x^2 + q^2) - {\omega}_M{{{\Omega}}}q_z^2 = 0
    \qwith {{{\Omega}}} = {\omega}_0 + i{\alpha}{\omega} + A_{\rm ex}(q_x^2 + q^2).
\label{eqn:qxj}
\end{equation}
%-----------------------------------------
and $a_j, b_j$ are six coefficients to be determined by the six boundary conditions in \Eq{eqn:bc-psi}. By \Eqs{eqn:psi}{eqn:phix}, the six boundary conditions \Eq{eqn:bc-psi} can be transformed into a set of linear equations with six unknowns $a_j, b_j$:  
%-----------------------------------------
\begin{equation}
    M(\qq, {\omega}) \smatrix{a_j \\ b_j} = 0
\label{eqn:Mab}
\end{equation}
%-----------------------------------------
where $M(\qq, {\omega})$ is a $6{\times}6$ matrix depending on the material parameters and injected spin current: ${\omega}_0, {\alpha}, k_s, k_j$. The dipolar-exchange spin wave dispersion is determined by the condition that the determinant of the coefficient matrix vanishes: $\abs{M(\qq, {\omega})} = 0 \Ra {\omega}(\qq)$, whose real part represents the energy and imaginary part the inverse lifetime of the spin wave mode. The corresponding solution of \Eq{eqn:Mab} for $a_j, b_j$ gives the spin wave amplitude profile according to \Eq{eqn:psi}. 

The spin wave dispersion in magnetic thin films has been studied long ago \cite{de_wames_dipole-exchange_1970, hillebrands_spin-wave_1990, kalinikos_theory_1986}, but without current-induced spin-transfer torque  \cite{xiao_spin-wave_2012}, spin pumping, and/or surface anisotropy. Similar to the Gilbert damping, both the current-induced spin-transfer torque and spin pumping mainly affect the dissipative behavior via the imaginary part of the frequency, while the real part remains practically unchanged. While the surface anisotropy affects the spin wave dispersion itself or the real part of the frequency. From the definition of ${\psi}$ in \Eq{eqn:psi}, we see that the spin wave is amplified when $\imm{{\omega}(\qq)} < 0$, which indicates instability and $\imm{{\omega}(\qq)} = 0$ will be used as criteria for spin wave excitation with wave vector $\qq$.  

\begin{table}[t]
        \centering
        \begin{tabular}{c|l|l|c|l|l} \hline
        Parameter               & YIG                   & Unit          & 
        Parameter               & YIG                   & Unit          \\ \hline  
       $M_s$                   & $^a1.56{\times}10^5$          & A/m           &    
        ${\omega}_0 = {\gamma}H_0$      & $0.5{\omega}_M$                   & GHz                \\          
         ${\alpha}$                          & $^a6.7{\times}10^{-5}$    & -               & 
        ${\omega}_M={\gamma}{\mu}_0M_s$ & $34.5$                    & GHz                 \\          
         $g_r$                   & $^b3{\times}10^{17}$ &1/m$^2$    &
        $A_{ex}$                & $4.74{\times}10^{-6}$ & m$^2$/s       \\      
        $K_s$                   & $^c5{\times}10^{-5}$          & J/m$^2$       &
        $k_s=2{\gamma}K_s/A_{\rm ex}M_s$                        & $2.5{\times}10^7$             & 1/m              
          \\\hline          
        \end{tabular}
        \caption{Parameters for YIG.
                $^a$Ref. \cite{kajiwara_transmission_2010},
                $^b$$g_r$ = $10^{16}~{\sim}~10^{19}$/m$^2$, Ref. \cite{kajiwara_transmission_2010,Burrowes2012,czeschka_scaling_2011},             
                $^cK_s$ = $0.01 {\sim} 0.1$ erg/cm$^2$ or $10^{-5} {\sim} 10^{-4}$ J/m$^2$, Ref.
\cite{yen_magnetic-surface_1979, ramer_effects_1976}.
%               $^b$Ref. \cite{borghese_damped_1980},
                }
        \label{tab:param}
\end{table}

%=========================================================================
\section{Dispersion, amplification, and dissipation of spin waves in magnetic insulators} 

%In order to extend the simple analysis about the EASA induced surface spin wave above, 

We study in this section the dipole-exchange spin wave dispersion in a thin film of finite thickness. The full spin wave dispersion is anisotropic in $\qq$, multiple volume transverse modes exist due to the confinement in the $x$-direction, and magnetostatic surface modes due to the dipolar interactions exist. A proper description contains the following ingredients: finite thickness ($d = 0.61{\upmu}$m), intrinsic magnetic damping, exchange coupling, and dipolar fields in the bulk, and surface anisotropy, spin current injection, spin pumping at the interface. We calculate numerically the complex eigenfrequencies ${\omega}(\qq, k_j)$ as a function of the in-plane wave vector $\qq$ and the injected spin current $k_j$. $\im{{\omega}}$, the effective dissipation, can be either positive (damping) or negative (amplification) when driven by the spin-transfer torque. In order to understand the consequences of surface anisotropy, we study three different cases: i)  free surface without surface anisotropy, ii) easy-axis surface anisotropy (EASA), and iii) hard-axis (easy-plane) surface anisotropy (HASA). We consider  spin pumping in Section \ref{spinpumping} and \ref{excitationpower}, but initially assume $k_{p}=0$ (zero spin pumping). 

%****************************************************************
\subsection{Without surface anisotropy} 

%-----------------------------------------
\begin{figure}[t]
        \includegraphics[width=\textwidth]{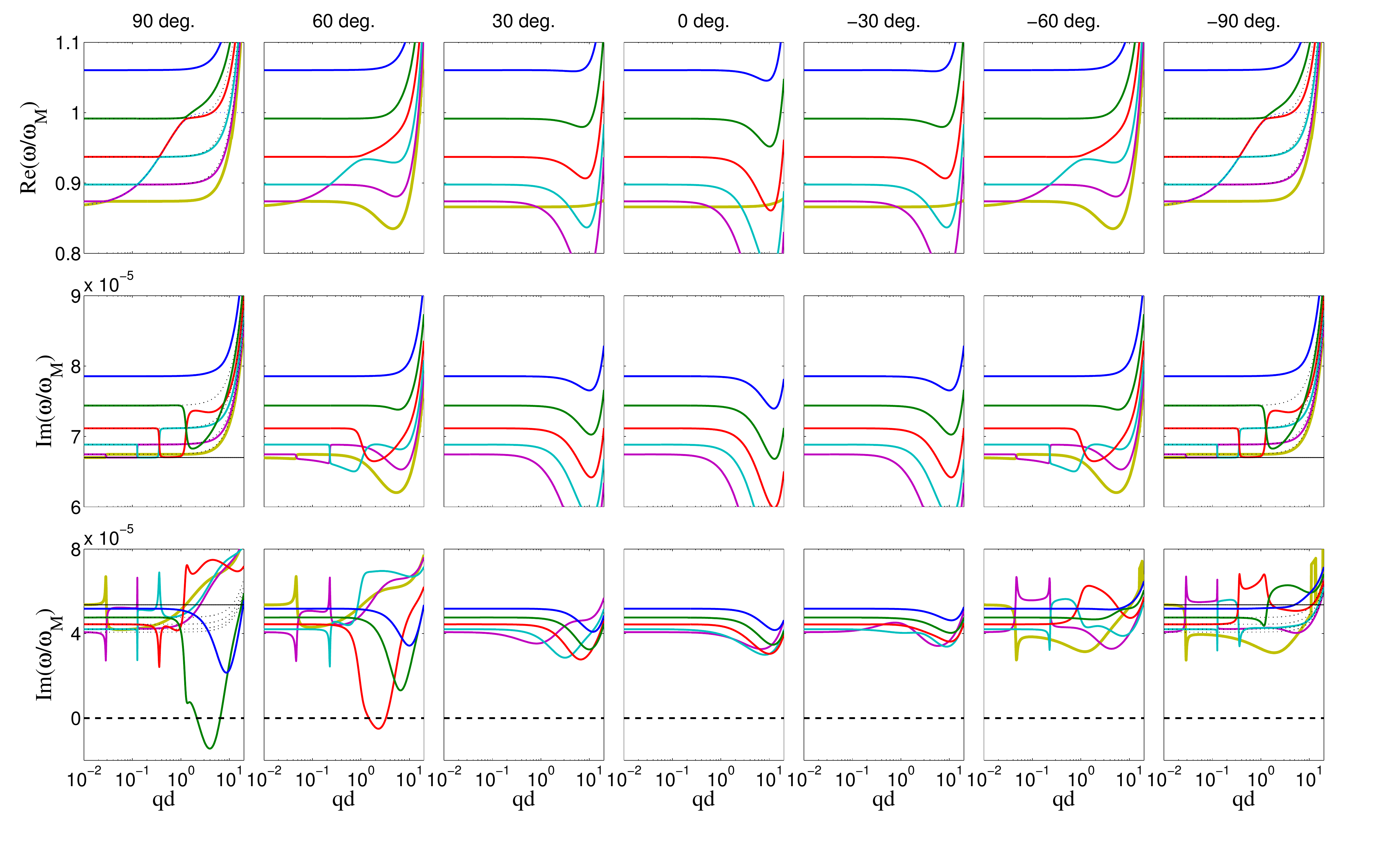}
    \caption{Spin wave dispersion and dissipation at various angle ${\theta}$ for $k_s = 0$. Top: $\re{{\omega}}$, middle: $\im{{\omega}}$ at $k_j = 0$, bottom: $\im{{\omega}}$ at $k_j = -0.2k_c$. }
        \label{fig:ks0a}
\end{figure}
%-----------------------------------------

%Dispersion without surface anisotropy.
First, we disregard the surface anisotropy: $k_s = K_s = 0$. The solution of \Eq{eqn:Mab} gives the dispersion and dissipation versus the in-plane wave vector $\qq$ as shown in \Figure{fig:ks0a} for different angles ${\theta} = {\angle}(\mm,\qq)$ between $\qq$ and the equilibrium $\mm$ (or $\hzz$). Without spin current injection, the dispersion (top row) and dissipation (middle row) are the same for ${\theta}$ and $-{\theta}$ because the structure is axially symmetric around the $\hzz$-axis. However, the spin current injected on the top surface breaks this symmetry, and the dissipation curves (bottom row) differ for ${\theta}$ and $-{\theta}$, while the dispersion remains more or less symmetric because the spin current injection affects the dissipation to first order and dispersion to  second order  in $k_j$.

%-----------------------------------------
\begin{figure}[t]
        \includegraphics[width=\textwidth]{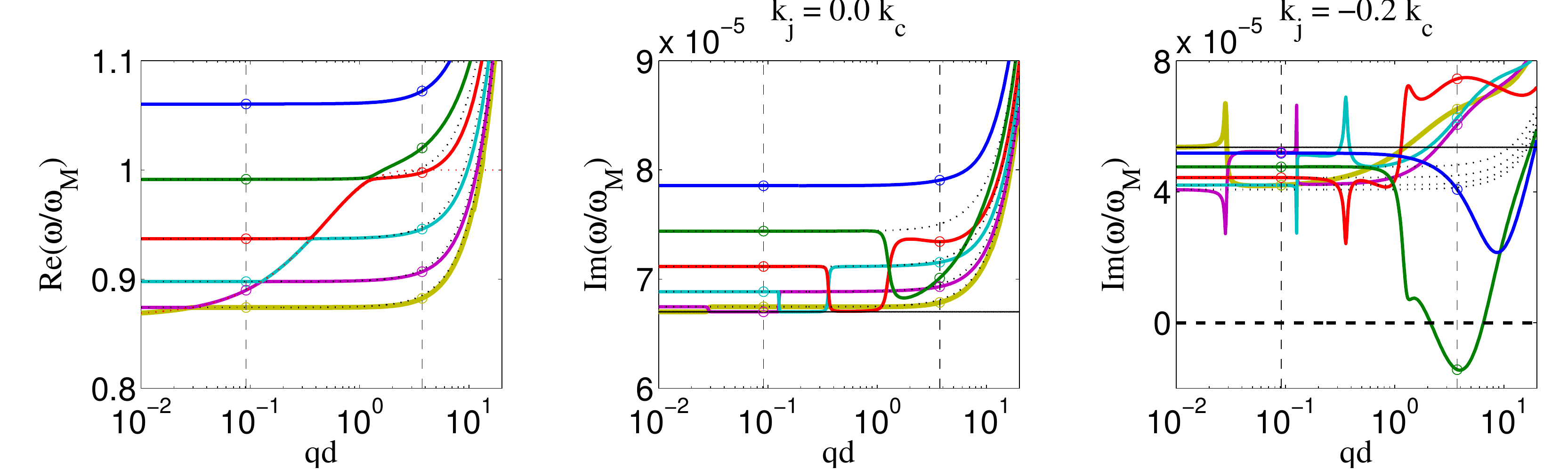}
    \caption{Spin wave band structure in YIG for $d = 0.61~\upmu$m without surface anisotropy ($k_s = 0$) at ${\theta} = {\angle}(\mm,\qq) = {\pi}/2$. From left to right: $\re{{\omega}/{\omega}_M}$, $\im{{\omega}/{\omega}_M}$ at $k_j = 0$, $\im{{\omega}/{\omega}_M}$ at $k_j = -0.2k_c$. The dashed curves are plotted using the analytical expression for ${\omega}$ in \Eq{eqn:wnwMSW}. }
        \label{fig:ks0}
\end{figure}
%-----------------------------------------

Let us focus on the results for ${\theta} = {\pi}/2$ in \Figure{fig:ks0}. The left and middle panels show the dispersion $\re{{\omega}}$ and dissipation $\im{{\omega}}$ without spin current injection. In the plot of $\re{{\omega}}$, the magnetostatic surface wave (MSW) crosses the bulk modes \cite{de_wames_dipole-exchange_1970}. The dispersion for the $n$th bulk mode and the (independent) MSW for ${\theta} = {\pi}/2$ and ${\alpha}~{\ll}~1$ is given by 
%-----------------------------------------
\begin{subequations}
\label{eqn:wnwMSW}
\begin{align}
    {\omega}_n &= \sqrt{ {\omega}_{nq} ({\omega}_{nq} + {\omega}_M) }
    +i\midb{\left({\alpha}+2{A_{ex}k_p\ov{\omega}_0d}\right)\smlb{{\omega}_{nq} + {{\omega}_M\ov 2}} + 2A_{\rm ex}{k_j\ov d}}
    \qwith n = 1, 2, {\dots}
%    \qwith {\omega}_{nq} = {\omega}_0 + A_{\rm ex}\midb{\smlb{n{\pi}\ov d}^2 + q^2}
    \label{eqn:wn} \\
    {\omega}_{\rm MSW} &= \sqrt{\smlb{{\omega}_0 + {{\omega}_M\ov 2}}^2- {{\omega}_M^2\ov 4}e^{-2qd}}
    + i\midb{\left({\alpha}+{A_{ex}k_p\ov{\omega}_0d}\right)\smlb{{\omega}_0+{{\omega}_M\ov 2}} + A_{\rm ex}{k_j\ov d}}
    \qfor qd < 1. \label{eqn:wMSW}
\end{align}
\end{subequations}
%-----------------------------------------
with ${\omega}_{nq} = {\omega}_0 + A_{\rm ex}[q^2 + (n{\pi}/d)^2]$.   

\begin{figure}[t]
        \includegraphics[width=\textwidth]{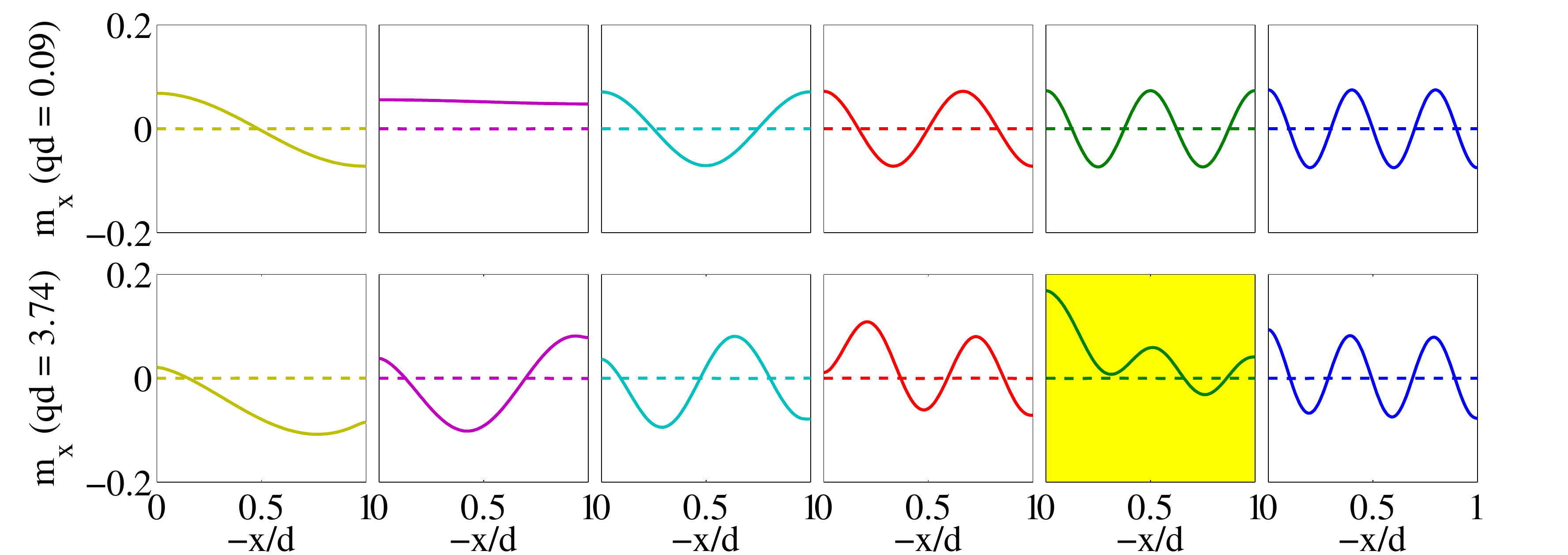}
    \caption{$m_x$ of the same 6 modes for $qd = 0.09$ and $3.74$ indicated by the dashed vertical lines in \Figure{fig:ks0}. The colors label different bands. The amplitude of the green mode mode (shaded/yellow panel) is amplified.}
        \label{fig:ks0profile}
\end{figure}

When spin current is injected, $\im{{\omega}}$ decreases/increases depending on the polarity of the injected spin current, as expected. For negative $k_j$, dissipation decreases for all modes as seen in the right panel of \Figure{fig:ks0}. In the regime $qd~<~1$, the bulk modes become less damped than the MSW mode as indicated in \Eq{eqn:wnwMSW}, which is counter intuitive but can be explained qualitatively using the spin wave profiles in \Figure{fig:ks0profile}. Here the spin wave amplitude profiles are plotted for the modes indicated by the vertical dashed line in \Figure{fig:ks0} at two  values  $qd = 0.09, 3.74$. Similar to $\re{{\omega}}$ the amplitudes  are only weakly affected by the spin current injection. The top row in \Figure{fig:ks0profile} is for small $qd = 0.09~ {\ll}~ 1$, thus \Eq{eqn:wMSW} is valid in this regime. The purple (2nd) and the cyan (3rd) curves in the top panels in \Figure{fig:ks0profile}  show the MSW and the second bulk mode respectively. Though the purple MSW is called a ``surface'' wave, it is not really localized at the surface because of its small $q: 1/q~{\sim}~d/0.09~{\gg} ~d$, therefore it is close to a uniform mode. The cyan bulk mode is actually ``lighter'' than the purple MSW mode in terms of total magnetization because of its oscillating magnetization. The dissipation decrease/increase due to the spin current injection is proportional to the boundary value of $\mm_{\perp}(0)$ and inversely proportional to the total magnetization of the mode; therefore $\im{{\omega}}$ of the bulk mode decreases/increases more strongly here than the MSW mode under spin current injection.

%-----------------------------------------
%-----------------------------------------

The current-induced dissipation in \Eq{eqn:wnwMSW} is no longer valid for the modes in the bottom panels in \Figure{fig:ks0profile} at $qd = 3.74 > 1$, at which the MSW and the bulk mode hybridize as seen in the red (4th) and green (5th) amplitude profile in the lower panels. Both modes have surface and bulk features. From their profiles, we can infer that the dissipation for the red mode should be weakly influenced by the spin current injection because of its nearly vanishing $\mm_{\perp}(0)$, whereas the green mode should be strongly affected because of its large $\mm_{\perp}(0)$. Indeed, in the right panel in \Figure{fig:ks0}, the dissipative part $\im{{\omega}}$ for the green mode dips into the negative regime, \ie is amplified, while the other modes are still damped.     

The spiky features in the right-most panel of \Figure{fig:ks0} are not  artifacts: they are caused by mode anti-crossings, at which the two modes form bonding and anti-bonding states. The spin wave profiles $\mm(x)$ for the two modes add up (subtract) to form the (anti-)bonding  at the anti-crossing point, therefore $\mm_{\perp}(0)$ for the two modes also add up (subtract). As the mode transforms from bulk mode to MSW or vice versa when being tuned through the (anti-)crossing, $\mm_{\perp}(0)$ decreases/increases first then increases/decreases again. Precisely at the anti-crossing point one mode has a maximized $\mm_{\perp}(0)$ while  that of the other mode is minimal. For example, at the point where the red mode has a dip and the cyan mode has a peak in \Figure{fig:ks0}, $\mm_{\perp}(0)$ for the red mode is large while $\mm_{\perp}(0)$ for the cyan mode is negligibly small. Therefore, upon the application of spin current, the dissipation for the red mode decreases and the cyan mode is not affected (and is very close to its value for $k_j = 0$). 

By its different chirality, an analogous hybrid MSW at ${\theta} = -{\pi}/2$ is localized at the opposite surface to vacuum ($x = -d$), so $\mm_{\perp}(0)$ is small. This means the spin current induced torque at $x =0$ is small and its dissipation is only weakly affected by the spin current injection.

%****************************************************************
\subsection{With easy-axis surface anisotropy} 

\begin{figure}[t]
        \includegraphics[width=\textwidth]{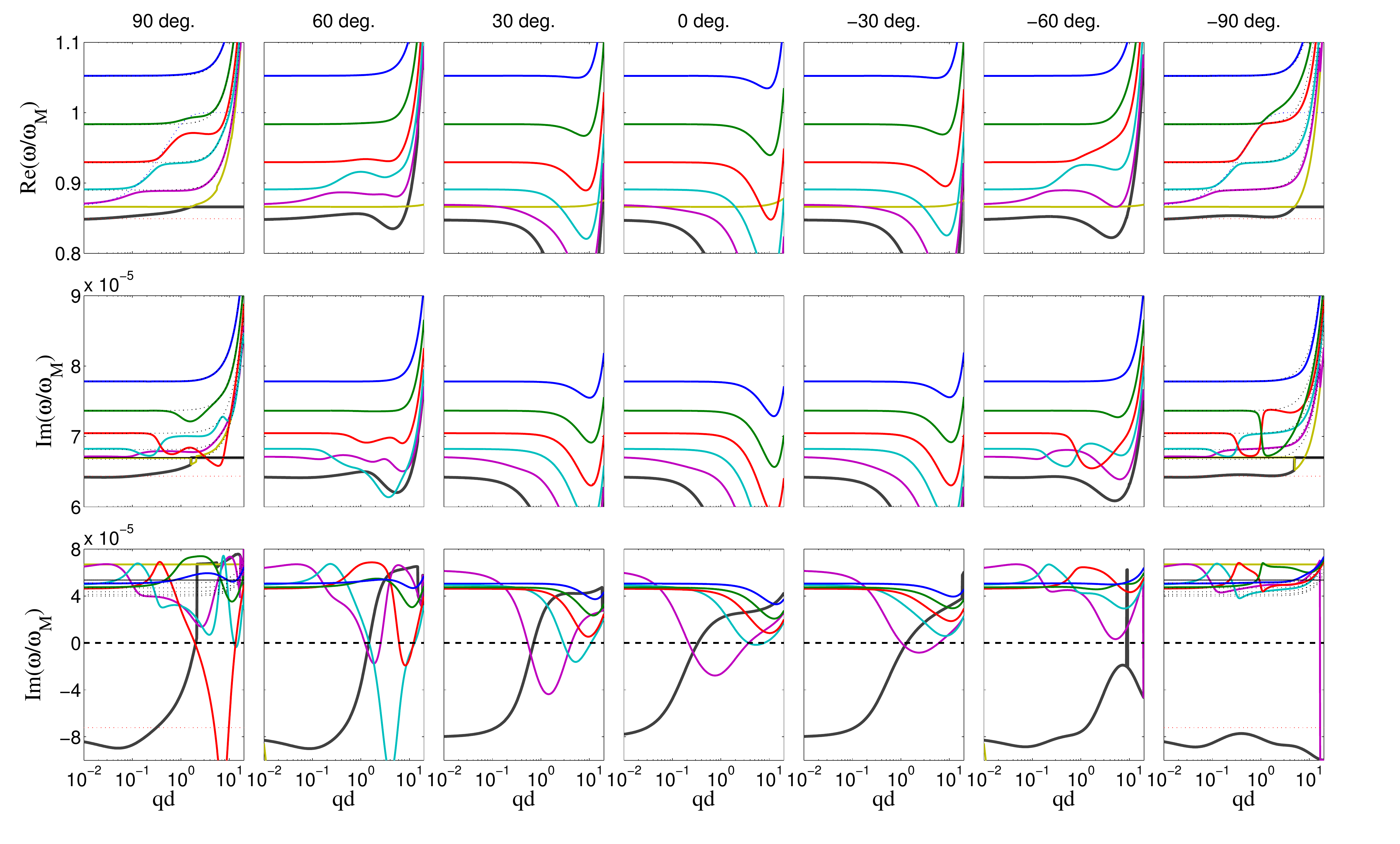}
    \caption{Spin wave dispersion and dissipation at various angle ${\theta}$ for $k_s = 25/\upmu$m. Top: $\re{{\omega}}$, middle: $\im{{\omega}}$ at $k_j = 0$, bottom: $\im{{\omega}}$ at $k_j = -0.2k_c$. }
        \label{fig:ks1a}
\end{figure}

We now include the effects of an easy-axis surface anisotropy (EASA) at the top surface. For $k_s > 0$ the anisotropy energy is minimized when the magnetization is parallel to the surface normal. We assume that  the anisotropy is not strong enough to enforce a perpendicular equilibrium magnetization. The solution of \Eq{eqn:Mab} gives the dispersion and dissipation versus the in-plane wave-vector $\qq$ which for these conditions are given  in \Figure{fig:ks1a} for different angles ${\theta}$. Different from the $k_s = 0$ case in \Figure{fig:ks0a}, without spin current injection, the dispersion (top row) and dissipation (middle row) are no longer the same for ${\theta}$ and $-{\theta}$, because the surface anisotropy is on the top surface only and breaks the axial symmetry around the $\hzz$-axis. Comparing dispersion for ${\theta} = {\pi}/2$ and ${\theta}= -{\pi}/2$ in \Figure{fig:ks1a} and \Figure{fig:ks0a}, we find that the ${\theta} = {\pi}/2$ configuration is modified far more than the ${\theta} = {-{\pi}}/2$ one by the surface anisotropy. At ${\theta} = {\pi}/2$ case, both MSW and the EASA induce surface spin waves (discussed below) that are localized at the top surface. They are therefore strongly coupled. However, for the ${\theta} = {-{\pi}}/2$ case, the MSW is localized at the bottom surface, far away from the EASA induced surface spin wave. The spatial separation obviously reduces the coupling and the individual bands remains more or less intact.  

%-----------------------------------------
%-----------------------------------------

\begin{figure}[t]
        \includegraphics[width=\textwidth]{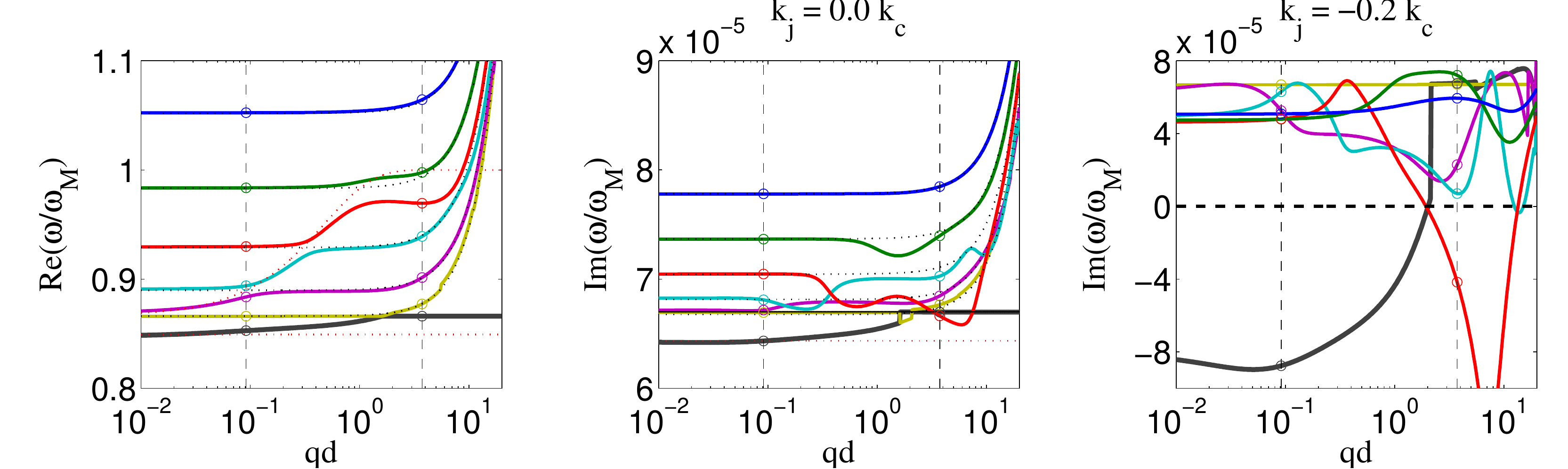}
    \caption{Same as \Figure{fig:ks0} but with $k_s = 25/\upmu$m. The black (red) dashed curves are plotted using the analytical expression in \Eq{eqn:wnks} (\Eqs{eqn:wS}{eqn:wSi}) for the bulk modes (EASA surface wave) at zero spin current.} 
        \label{fig:ks1}
\end{figure}

%Dispersion with surface anisotropy.
Let us focus now on ${\theta} = {\pi}/2$ in order to study the effect of surface anisotropy ($k_s > 0$ for EASA) on the dispersion and dissipation of the spin waves. In the absence of spin current injection ($k_j = 0$), the surface anisotropy shifts the eigenfrequency of the $n$-th bulk mode to (assuming $A_{\rm ex}k_s^2~{\ll}~{\omega}_0$ and $qd~{\lesssim}~1$): 
%-----------------------------------------
\begin{equation}
    {\omega}'_n = {\omega}_n -
    {A_{\rm ex}k_s\ov d}
    \midb{{{\omega}_{nq}\ov \re{{\omega}_n}}
%    {(2{\omega}_{nq}+{\omega}_M)\sqrt{({\omega}_{nq}+{\omega}_0+{\omega}_M)}\ov
%     (2{\omega}_{nq}+{\omega}_M)\sqrt{({\omega}_{nq}+{\omega}_0+{\omega}_M)} - ({\omega}_{nq}+{\omega}_M)\sqrt{A_{\rm ex}k_s^2} }
    \smlb{ 1 - {{\omega}_{nq}+{\omega}_M\ov 2{\omega}_{nq}+{\omega}_M}
    \sqrt{A_{\rm ex}k_s^2\ov {\omega}_{nq}+{\omega}_0+{\omega}_M} }^{-1}
    +i{\alpha}},
%    {\omega}_{\rm EASA} &= \sqrt{{\omega}_0({\omega}_0+{\omega}_M)}
%    \bigb{1 -
%    {[(k_s+2q){\omega}_0+q{\omega}_M]^2\ov 
%    2{\omega}_0({\omega}_0+{\omega}_M)\midb{(k_s+2q)+\sqrt{2{\omega}_0+{\omega}_M\ov A_{\rm ex}}}^2}
%    }
\label{eqn:wnks}
\end{equation}
%-----------------------------------------
where ${\omega}_n$ is given by \Eq{eqn:wn} for $k_s = 0$. Both frequency and dissipation decrease with increasing surface anisotropy $k_s$. We may compare \Eq{eqn:wnks} (thin dashed line) with the numerical calculations (thick solid line) in \Figure{fig:ks1}. At zero spin current injection (the left and middle panels), both real and imaginary parts are well represented by \Eq{eqn:wnks} in the present parameter regime. 

%-----------------------------------------
%-----------------------------------------

%****************************************************************
%\subsubsection{The role of surface anisotropy} 

Apart from the eigenfrequency shifts of the bulk modes, the main differences between $k_s = 0$ (\Figure{fig:ks0}) and $k_s > 0$ (\Figure{fig:ks1}) cases are: i) appearance of an additional (black) band, whose frequency (at $q = 0$) is less than the lowest macrospin mode frequency $\sqrt{{\omega}_0({\omega}_0+{\omega}_M)}$, ii) strongly modified spin wave hybridization for $qd ~{\gtrsim} ~1$ by the new band in i). 

In order to understand the differences caused by EASA, we study the limit $d\ra {\infty}$, \ie the magnetic film is semi-infinite and $b_j = 0$ in \Eq{eqn:phix}. Focusing for simplicity on vanishing in-plane wave-vector $\qq = (q^y, q^z) = 0$, \Eq{eqn:phi-doe} becomes of 4th-order and the scalar potential can be written as: 
%-----------------------------------------
\begin{equation}
    {\psi}(\rr) = {\sum}_{j=1}^2a_je^{iq_jx}e^{i{\omega}t} \qwith
    q_j({\omega}) = -i\smlb{{\omega}_0+\half {\omega}_M {\pm} \sqrt{{\omega}^2+\quarter{\omega}_M^2} {\pm} i{\alpha}{\omega}\over A_{\rm ex}}^{\half},
    \label{eqn:psis}
\end{equation}
%-----------------------------------------
where $q_{1,2}$ are negatively imaginary with $\abs{q_1}{\gg} \abs{q_2}$. Imposing the boundary conditions from \Eq{eqn:bc-psi} at $x = 0$,  $\abs{M(\qq, {\omega})} = 0$ leads to (up to the first order in $k_j$):
%-----------------------------------------
\begin{equation}
        2q_1q_2(q_1+q_2) +ik_s\midb{(q_1+q_2)^2+{{\omega}_M\ov A_{\rm ex}}} + 4k_j{\omega} = 0.
%       q_1q_2(q_1+q_2) -ik_s\midb{q^2-q_1q_2+{{\omega}_0\ov A_{\rm ex}}} + 2k_j{\omega} = 0.
\label{eqn:det}
\end{equation}
%-----------------------------------------
The solutions of \Eq{eqn:det} are the {\it complex} eigenfrequencies ${\omega}$. 

To 0th-order in dissipation, \ie with vanishing bulk damping (${\alpha} = 0$) and spin current injection ($k_j = 0$), \Eq{eqn:det} simplifies to $k_s = 2iq_2/[1+{\omega}_M/(A_{\rm ex}q_1^2)]$ using $\abs{q_1} {\gg} \abs{q_2}$. There is no real solution when surface normal is the hard axis or $k_s \le 0$. However, when the surface normal is an easy axis or $k_s > 0$, there is a single real solution ${\omega}_S$. For ${\eta}_s {\equiv} \sqrt{A_{\rm ex}k_s^2/(2{\omega}_0+{\omega}_M)}~{\ll}~1$
%-----------------------------------------
\begin{align}
    \re{{\omega}_{\rm S}}~&{\simeq}~\sqrt{{\omega}_0({\omega}_0+{\omega}_M)}
    \midb{1 - {\eta}_s {{\omega}_0 \ov 2({\omega}_0+{\omega}_M)} - {\eta}_s^2 {{\omega}_0\ov 2{\omega}_0+{\omega}_M}  }, \label{eqn:wS} \\
%    - {2{\omega}_0^2A_{\rm ex}k_s^2\ov (2{\omega}_0+{\omega}_M)\sqrt{{\omega}_0({\omega}_0+{\omega}_M)}}, \\
%    < \sqrt{{\omega}_0({\omega}_0+{\omega}_M)}, \\
    q_1~&{\simeq}~-i\sqrt{2{\omega}_0+{\omega}_M\ov A_{\rm ex}} \qand
    q_2~{\simeq}~-i\sqrt{{\omega}_0^2/(2{\omega}_0+{\omega}_M)\ov A_{\rm ex}} \smlb{{\eta}_s + {\eta}_s^2{{\omega}_0+{\omega}_M\ov 2{\omega}_0+{\omega}_M}}. 
\nonumber
\end{align}
%-----------------------------------------
Since $q_{1,2}$ are both negative imaginary, the corresponding spin waves in \Eq{eqn:psis} are  localized near the surface. Because $\abs{q_1}~{\gg}~\abs{q_2}$, the penetration depth of a surface wave is roughly $d_s = 1/\abs{q_2}~ {\propto}~ 1/k_s$, which proves the direct involvement of the surface anisotropy. Note that, in contrast to the  MSW, the EASA-induced surface spin wave is not chiral, \ie its amplitude profiles are identical for $\qq$ and $-\qq$ (if not hybridized with the MSW).  

To the 1st-order in the dissipation, \ie  in ${\alpha}$ and $k_j$, we find 
%-----------------------------------------
\begin{align}
\im{{\omega}_{\rm S}} &= i\sqrt{\re{{\omega}_{\rm S}}^2+{{\omega}_M^2\ov 4}} 
\midb{1 - {\eta}_s^2{2{\omega}_0{\omega}_M \ov (2{\omega}_0+{\omega}_M)^2} 
\smlb{1 + {\eta}_s{2{\omega}_M^2 - {\omega}_0^2\ov {\omega}_M^2+2{\omega}_0{\omega}_M}}}
\label{eqn:wSi} \\
&{\times}
\bigb{{\alpha} + {\eta}_s\sqrt{A_{\rm ex}\ov 2{\omega}_0+{\omega}_M}
\midb{
k_j {2{\omega}_0\ov 2{\omega}_0+{\omega}_M} \smlb{1+{\eta}_s{{\omega}_0+2{\omega}_M\ov 2{\omega}_0+{\omega}_M}}
+k_p \smlb{1+{\eta}_s{{\omega}_0+{\omega}_M\ov 2{\omega}_0+{\omega}_M}} }} . \nonumber
\end{align}
%-----------------------------------------
The eigenfrequencies in \Eqs{eqn:wS}{eqn:wSi} are derived for $\qq = 0$ and for a semi-infinite film. However, we still expect similar surface spin wave for film thickness $d~{\gg}~1/k_s$. For a YIG thin film with $d = 0.61~\upmu$m, using the parameters given in Table \ref{tab:param}, we estimate the penetration depth $d_s~ {\simeq}~ 1/\abs{q_2} = 0.13~\upmu$m ${\ll} ~d$, so \Eqs{eqn:wS}{eqn:wSi} still hold. The frequency and dissipation given in \Eqs{eqn:wS}{eqn:wSi} are shown as the red dashed horizontal line in \Figure{fig:ks1}, which agrees very well with the black band in the same figure at the point $q = 0$.    

According to \Eq{eqn:wSi} the imaginary part of the eigenfrequency for the surface mode becomes negative when
%-----------------------------------------
\begin{equation}
        k_j < k_j^c = -{{\alpha}\ov k_s}{ (2{\omega}_0+{\omega}_M)^2\ov 4A_{\rm ex}{\omega}_0} 
        + {\alpha} {{\omega}_0+2{\omega}_M\ov 4{\omega}_0}\sqrt{2{\omega}_0+{\omega}_M\ov A_{\rm ex}}+{{2{\omega}}_0+{\omega}_M\ov 2{\omega}_0}k_{p},
\label{eqn:kjc}
\end{equation}
%-----------------------------------------
where $k_j^c$ is the threshold current for exciting the EASA induced surface spin wave. When $A_{\rm ex}k_s^2~{\ll} ~{\omega}_0$, the first term in \Eq{eqn:kjc} dominates and the threshold current is proportional to the penetration depth of the surface spin wave, $d_s {\sim} 1/k_s$, because the threshold current is directly proportional to the total magnetization that is excited. Using the parameters in Table \ref{tab:param}, the threshold current for exciting the EASA induced surface spin wave (at $\qq = 0$) is $k_j^c = -0.08k_c$.  

The spin current injection has little effect on the dissipation of MSW because the MSW is only well defined for $qd~{\lesssim}~1$ and its decay length is comparable to or larger than the thickness of the film $d$, which means that a MSW at small wave vectors is similar to the bulk modes. On the other hand, the EASA-induced surface spin wave decays with $d_s{\sim} 1/k_s$ and is strongly localized near the surface for sufficiently large $k_s$. Therefore the spin current injection can strongly affect this mode because of its small total magnetization \cite{sandweg_enhancement_2010}. As seen in the right panel of \Figure{fig:ks1}, almost the whole black band is strongly amplified by a spin current injection of $k_j = -0.2k_c$. 

%-----------------------------------------
\begin{figure}[t]
        \includegraphics[width=\textwidth]{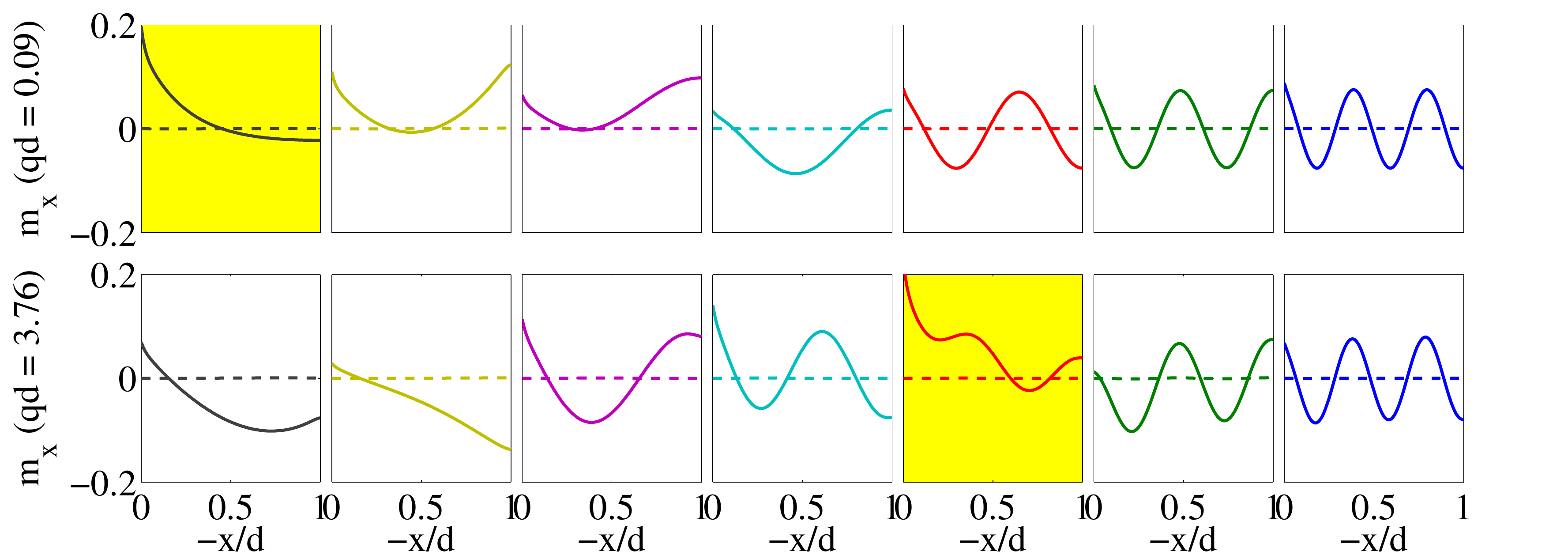}
        \caption{Same as \Figure{fig:ks0profile} but with $k_s = 25/\upmu$m.} 
        \label{fig:ks1profile}
\end{figure}
%-----------------------------------------

Inspecting the spin wave profiles in \Figure{fig:ks1profile} at two different $q$ values, we observe a surface spin wave near $x = 0$ for the black band at small $q$ (top yellow panel). At larger $q$ (bottom panels), the 1st (black) band loses its surface wave features to the 5th (red) band (see right panel in \Figure{fig:ks1}). The red band mode starts out as a MSW, but the EASA enhances its surface localization by hybridization with the black mode to become strongly amplified by the spin current at higher $q$, which is seen in the lower yellow panel of \Figure{fig:ks1profile}.

%-----------------------------------------
\begin{figure}[t]
        \includegraphics[width=\textwidth]{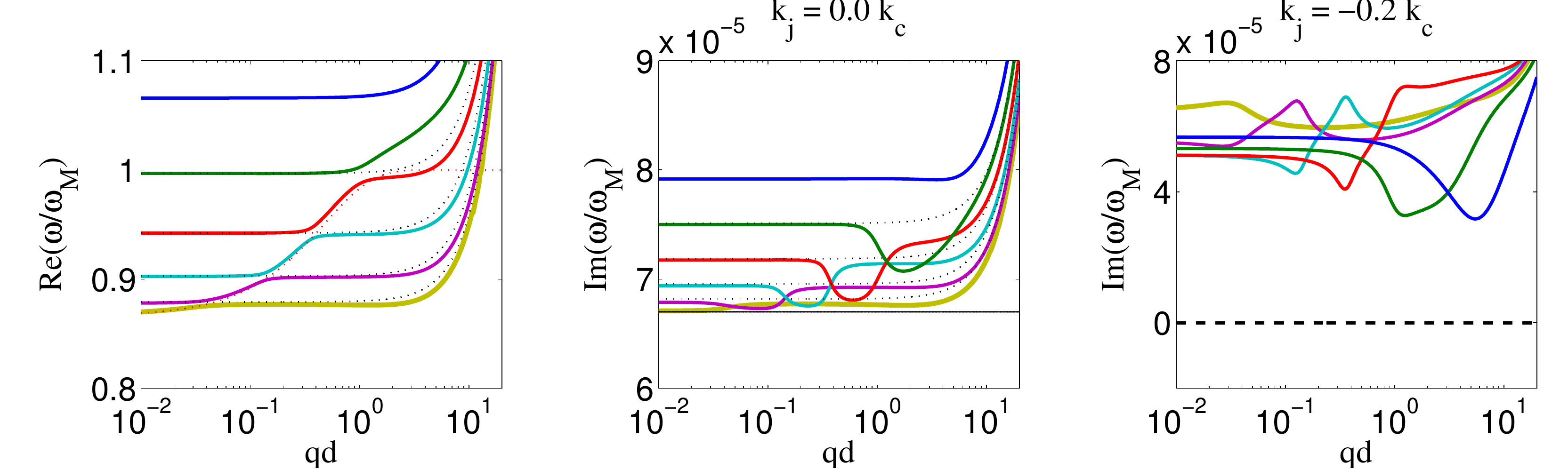}
    \caption{Same as \Figure{fig:ks0} but with $k_s = -25/\upmu$m. The black dotted curves are plotted using \Eq{eqn:wnks}.} 
        \label{fig:ks2}
\end{figure}
%-----------------------------------------
%

%****************************************************************
\subsection{With hard-axis surface anisotropy} 

As mentioned in the previous subsection, there is no surface mode solution to \Eq{eqn:det}  in the presence of a hard-axis surface anisotropy ($k_s < 0$) favoring interface spins that lie in plane. The dispersion for $k_s < 0$ shown in \Figure{fig:ks2}  is similar to that for $k_s = 0$ except for a shift of the bulk modes as expected from \Eq{eqn:wnks}. However, we observe that the spin current has little effect now since all modes avoid the interface such that $\mm_{\perp}(0)$ is small.  

%****************************************************************
\subsection{Excitation power spectrum} 
\label{excitationpower}

We introduce an approximate power spectrum that summarizes the information about the mode-dependent current-induced amplification as a sum over bands with band index $n$:
%-----------------------------------------
\begin{equation}
        P({\omega}) = {\sum}_n{\int}_{\rm \im{{\omega}_{\it n}}<0} \abs{\rm \im{{\omega}_{\it n}(\qq)}} 
        {\delta}[{\omega}-\re{{\omega}_{\it n}(\qq)}]d\qq,
\label{eqn:Pw}
\end{equation}
%-----------------------------------------
which is the density of states at frequency ${\omega}$ weighted by its amplification. As a disclaimer, we note that \Eq{eqn:Pw} merely gives partial information about the excitation ignoring \eg  their spin pumping once excited. It fails to describe any self-organization due to the coupling between excited modes. Nevertheless, it possibly provides a qualitative picture of the power spectrum near the threshold. 

%\Figure{fig:Pw} compares the approximated power spectrum defined in \Eq{eqn:Pw} for different combination of surface anisotropy and spin pumping. In the following, we separate the discussions for the case with and without spin pumping. 

\begin{figure}[t]
\centering
        \includegraphics[width=\textwidth]{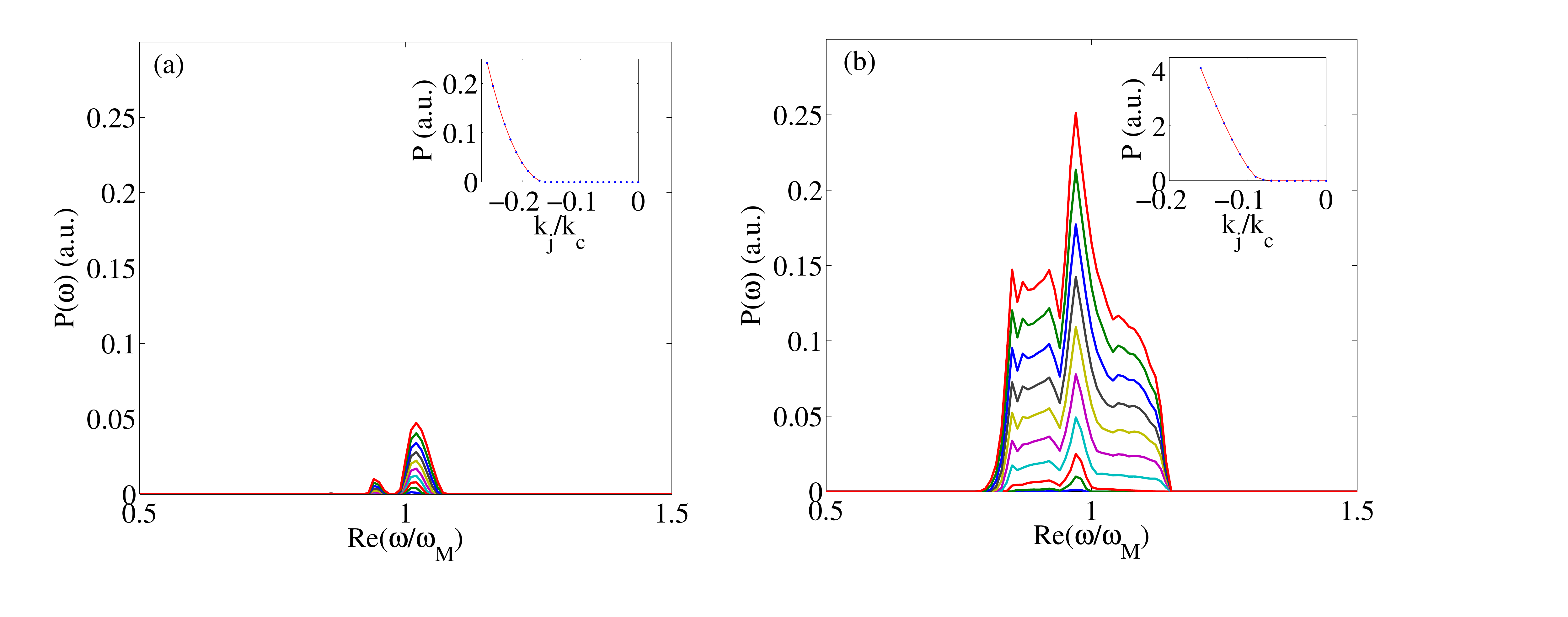}
%        \includegraphics[width=1.3\textwidth,trim= 50 0 0 0, clip=true]{fig/kpeffect_spectrum}
%         \caption{Power spectrum (resolution ${\delta}{\omega}/{\omega}_M = 0.01$) at various current levels ($k_j = -0.2k_c$ from the top decreasing by ${\delta}k_j = 0.01k_c$): (a)  $k_s = 0, k_p=0$; (b) $k_s = 25/{\upmu}$m,$ k_p=0$; (c) $k_s = 0, k_p=0.01/{\upmu}$m and (d) $k_s = 25/{\upmu}$m, $k_p=0.01/{\upmu}$m. Inset: the integrated power versus $k_j$.
% %Note $x$-axis is slightly different for the left and right panels, the $y$-scale is about 10-100 times smaller than the left panels.
%         } 
        \caption{Power spectrum (resolution ${\delta}{\omega}/{\omega}_M = 0.01$) at various current levels (decreasing by ${\delta}k_j = 0.01k_c$): (a)  $k_s = 0$, (b) $k_s = 25/{\upmu}$m;   We plot ten current levels exceeding the threshold critical current. Insets: the integrated power versus $k_j$.
        }
        \label{fig:Pw}
\end{figure}

 Without surface anisotropy (see  \Figure {fig:Pw} (a)), only a few modes are excited even at a relatively large current. However, when $k_s = 25/\upmu$m as shown in \Figure {fig:Pw} (b), the excitation is strongly enhanced by more than two orders of magnitude due to the easily excitable surface spin wave modes. Furthermore, we observe broadband excitation over a much larger range of frequencies.  This power spectrum is rather smooth, while the experiments by Kajiwara \etal  \cite{kajiwara_transmission_2010} show a large number of closely spaced peaks. The latter fine structure is caused by size quantization of spin waves due to the finite lateral extension of the sample that has not been taken into account in our theory since it complicates the calculations without introducing new physics. The envelope of the experimental power spectrum compares favorably with the present model calculations.
%This approximate power spectrum is continuous because the structure is assumed to be infinite in the latteral direction. When the latteral size is finite, the power spectrum becomes discretized and shall mimic the experimental power spectrum in Ref.  \cite{kajiwara_transmission_2010}, and the envelope of the discretized spectrum shall be the same as the continuous spectrum in \Figure{fig:Pw}.
The insets in \Figure{fig:Pw} show the integrated power and allow the following conclusions: 1) the excitation power is enhanced by at least two orders of magnitude by the EASA; 2) the critical current for magnetization dynamics is $k_j~{\sim}~0.08k_c$ for $k_s = 25/\upmu$m, which agrees very well with the estimates from \Eq{eqn:kjc}. This critical current is about one order of magnitude smaller than that for the bulk excitation ($k_c$), and about half of that for the case without surface anisotropy ($k_j = -0.16k_c$). 
%For $k_s = 25/\upmu$m, it corresponds to $J_c = 3{\times}10^{10}$A/m$^2$ for ${\theta}_H = 0.01$ \cite{mosendz_quantifying_2010} and $3.8{\times}10^9$A/m$^2$ for ${\theta}_H = 0.08$ \cite{ando_electric_2008, liu_spin-torque_2011}. 
These values are calculated for a film thickness of $d = 0.61~\upmu$m, but should not change much for $d = 1.3~\upmu$m corresponding to the experiment \cite{kajiwara_transmission_2010}, because the excited spin waves are localized at the interface. 
%Compared to the original estimate $J_c~{\sim}~10^{11{\sim}12}$A/m$^2 $, the critical current for a surface spin wave excitation is much closer to the experimental value of $J_c~{\sim}~10^9$A/m$^2$ \cite{kajiwara_transmission_2010} (although these experiments report a very inefficient spin wave absorption in contrast to the present model assumption).

%-----------------------------------------
%Power spectrum.
%-----------------------------------------

%****************************************************************
\subsection{Spin pumping}
\label{spinpumping}
%-----------------------------------------
\begin{figure}[h]
\centering
\includegraphics[width=0.9\textwidth]{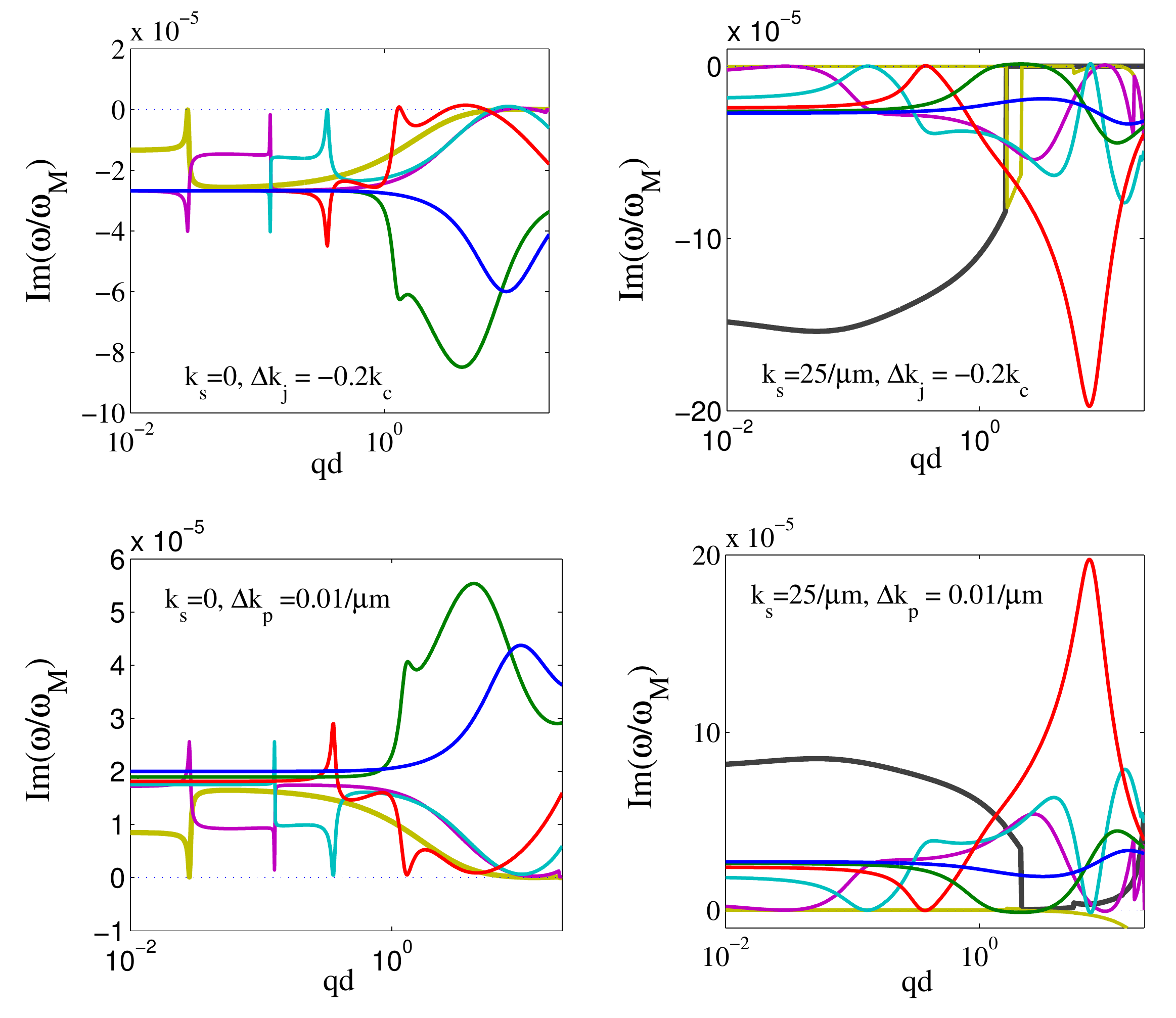}
\caption{Effects on spin wave dissipation by the spin transfer torque $k_j=-0.2k_c$ (the top two panels), obtained by subtracting the results for $k_j=0, k_p=0$ from that of $k_j=-0.2k_c, k_p=0$; the band shift due to spin pumping ${}k_p=0.01 /\upmu$m (the bottom two  panels), obtained by subtracting  the spin wave bands when $k_j=0, k_p=0$ from   the spin wave bands  with spin pumping only ($k_j=0,k_p=0.01/\upmu$m).}
\label{fig.spinband_kj_kp_diff} 
\end{figure}
%-----------------------------------------
The effect of spin pumping depends on the spin-mixing conductance at the interface, which also governs the spin-transfer torque. A complete theory  requires self-consistent treatment of both effects on an equal footing, which is still under study. In this section, we briefly discuss the  implications of spin pumping. 
% by assuming complete absorption of a transverse spin current at the Pt\textbar YIG interface \textit{i.e.} $ k_j$ directly represents the spin transfer torque
The values quoted for the spin mixing conductance vary strongly, \ie between $10^{16}~{\sim}~10^{20}$/m$^2$ \cite{kajiwara_transmission_2010,Burrowes2012,czeschka_scaling_2011,Rezende2013,jia_spin_2011}. A mixing conductance of  $g_r=3{\times}10^{17}/$m$^2$ as listed in Table \ref{tab:param} corresponds to $k_p=0.01/\upmu$m. This value is of the same order of magnitude as the Gilbert damping, \ie $k_p{\sim}{\alpha}{\omega}_0d/A_{ex}$ in \Eq{eqn:wn}. If $g_r$ is increased by one order of magnitude as predicted by first-principles calculations \cite{jia_spin_2011} and confirmed experimentally \cite{Burrowes2012,czeschka_scaling_2011}, the spin pumping will overwhelm the Gilbert damping,  but negligible when smaller by a factor of 10. 
%Thus the spin pumping is adopted at the same order of magnitude as that of spin transfer torque and Gilbert damping in this work. 
Similar to the Gilbert damping and spin-transfer torque, spin pumping mainly affects the imaginary part of the frequency and modifies the dissipation of the spin wave as in \Eqs{eqn:wnwMSW}{eqn:wSi}. \Eqs{eqn:wnwMSW}{eqn:wSi} also indicate that spin pumping enhances magnetic damping, and counteracts the spin-transfer torque for $k_j < 0$. 

In linear response, the effect of spin-transfer torque and spin pumping on the spin wave dissipation are simply additive and therefore plotted separately in \Figure{fig.spinband_kj_kp_diff}. The top two panels are the changes of the dissipation ${\delta}\im{{\omega}} = \im({\omega}_{k_j = -0.2k_c}) - \im({\omega}_{k_j=0})$ due to the spin-transfer torque  ($k_p = 0$), while the bottom two panels are ${\delta}\im{{\omega}} = \im({\omega}_{k_p = 0.01/{\upmu}m}) - \im({\omega}_{k_p=0})$ represent the effect of spin pumping only ($k_j = 0$). As both are proportional to the boundary value of magnetization, spin-transfer torque and spin pumping (for the chosen value of $k_p$) induce almost identical changes, but with opposite signs. The relative change for different modes can  be  understood by the boundary value of magnetization $\mm_{\perp}(0)$ observed in the spin wave profiles in \Figure{fig:ks0profile} and \Figure{fig:ks1profile}. Clearly, the MSW and EASA induced surface modes are more strongly affected by the spin pumping \cite{Kapelrud_2013}, thereby partly compensating the  reduced  critical current reported above for the same modes.  

%%-----------------------------------------
%\begin{figure}
%\centering
%%\includegraphics[width=1\textwidth,trim= 0 0 0 0, clip=true]{c:/dropbox/xz/spin_wave_exctation/figs/kpeffect_EASA.eps}
%\includegraphics[width=1\textwidth,trim= 0 0 0 0, clip=true]{fig/kpeffect_EASA}
%\caption{The same as Fig. \ref{fig.kpeffect_band_bulk} but with $k_s=25 {\upmu}m$.  From left to right: Re ${\omega}/{\omega}_M$, Im ${\omega}/{\omega}_M$, at $k_p=0$, Im ${\omega}/{\omega}_M$ at $k_p=0.01/{\upmu}m$.}
%\label{fig.kpeffect_band_EASA} 
%\end{figure}
%%-----------------------------------------

%=========================================================================
\section{Discussion} 

EASA enables a surface spin wave for the following reason: when $k_j = J_s = 0$, the boundary condition in \Eq{eqn:mbc} requires cancellation of the exchange and surface anisotropy torques: ${\partial}_xm_x - k_sm_x = {\partial}_xm_y = 0$ at $x = 0$. The exchange torque depends on the magnetization derivative in the normal direction, and can only take one sign in the whole film, and $m_{x,y} \ra 0$ as $x\ra -{\infty}$, therefore $(1/m_x){\partial}_x m_x > 0$. Torque cancellation (for a non-trivial solution) is therefore possible only for $k_s > 0$, which corresponds to an easy axis anisotropy (EASA). This EASA-induced surface spin wave only exists for the {\it in-plane} magnetized film ($m_z {\sim} 1$) as discussed here. For comparison, another type of surface spin waves exists for the {\it perpendicular} magnetized film ($m_x {\sim} 1$) with easy-plane surface anisotropy for $k_s < 0$, for which the boundary condition is ${\partial}_xm_{y,z}+k_sm_{y,z} = 0$ at $x = 0$
\cite{
%puszkarski_theoretically_1972, 
puszkarski_surface_1973,
%yu_tensorial_1974,
%yu_exchange-dominated_1975, 
wigen_microwave_1984, patton_magnetic_1984, kalinikos_theory_1986, gurevich_magnetization_1996}. 

According to \Eq{eqn:kjc}, critical current (excitation power) would be further reduced (increased) by a larger EASA. Ref. \cite{yen_magnetic-surface_1979} reports an enhancement of the YIG surface anisotropies for capped as compared to free surfaces. A Pt cover on a YIG surface \cite{kajiwara_transmission_2010} may enhance the surface anisotropy as well.  As seen from \Figure{fig:ks1}, the surface mode (black band) has group velocity ${\partial}{\omega}/{\partial}\qq$ comparable to that of the MSW. The excited surface spin wave therefore propagate and can be used to transmit spin information over long distance at a much lower energy cost than the bulk spin waves. On the other hand, they will also be more easily scattered by surface imperfections. 

The surface anisotropies of magnetic thin films with interface to vacuum or other materials are not well known. According to \Eq{eqn:wnks}, the bulk mode eigen-frequencies shift under the surface anisotropy. Brillouin light scattering experiments of the spin wave frequencies can therefore provide which gives direct information that can help determining both positive and negative $k_s$, unlike the EASA-induced surface wave, which only exists for $k_s > 0$. In the latter case, according to \Eq{eqn:wS}, it is also possible to determine the value of $k_s$ by measuring the eigen-frequency of the EASA-induced surface wave.

The EASA induced surface spin waves have the following properties, i) they can be easily induced (intentionally or unintentionally) by engineering the surface anisotropy, ii) its penetration depth is controllable by the strength of the surface anisotropy, iii) its excitation requires only small spin currents, iv) it has finite group velocity and can propagate long distance without much loss in the absence of surface roughness. The EASA induced surface spin wave is strongly localized at the surface, depending on the strength of EASA. This property means that the EASA surface waves are strongly damped by spin pumping, but only weakly absorb microwaves. Da Silva \etal indeed observed such behavior in a recent experiment \cite{da_silva_enhancement_2013}. Because the EASA surface wave only exists when there is EASA, it likely does not propagate naturally in uncovered sections of the film without EASA. In such circumstances, some intermode scattering would occur. This property could be used advantageously, \textit{e.g.} in order to fabricate surface spin wave guides on YIG films. If the complication due to the enhanced spin pumping can be controlled, this  type of surface spin wave mode might be superior  for spin information processing and transport. 

In the present study we were mainly concerned with the magnetization dynamics, disregarding the details of electron and spin transport in the normal metal, such as the spin current generation, spin accumulation at the interface \cite{tserkovnyak_enhanced_2002}. Although the values of the parameters  of spin current generation by the inverse spin Hall effect are still under debate, a more quantitative study of the transport processes in the normal metal is called for. 

%We did not include spin pumping that increases the dissipation of spin waves. Surface modes are expected to be affected more strongly than bulk modes. We estimate a spin pumping ${\propto} k_pm_{\perp}(0)$, where $k_p = ({\gamma}{\hbar}/A_{\rm ex}M_s)(g_r/4{\pi})$ with $g_r$ the real part of the mixing conductance. While the torque is proportional to the applied current, the pumping effect is more or less fixed up to a factor of order one determined by ratio between the mode frequency and ${\omega}_0$. This means that the spin pumping effect is important when $k_j {\lesssim} k_p$, \ie close to the excitation threshold, therefore it will increase the threshold current. This increase should be larger for the surface modes than the bulk modes, because the spin pumping does not distinguish surface or bulk modes as long as their $m_{\perp}(0)$ is the same. However, the quantitative influence on the threshold current for different modes and on the excitation spectrum is subject to further calculations.  

In conclusion, we predict that an easy-axis surface anisotropy gives rises to a surface spin wave mode, which reduces the threshold current required to excite the spin waves and dramatically increases the excitation power. Multiple spin wave modes can be excited simultaneously at different frequencies and wave vectors, thereby explaining recent experiments. Surface spin wave excitations could be useful in low-power future spintronics-magnonics hybrid circuits.

%=========================================================================
\section{Acknowledgment} 

This work was supported by the National Natural Science Foundation of China (No.  11004036, No. 91121002), the special funds for the Major State Basic Research Project of China (No. 2011CB925601),  the University Grant Council (AoE/P-04/08) of the government of HKSAR, the FOM foundation, DFG Priority Program SpinCat, EG-STREP MACALO, the ICC-IMR, and ReiMei project.

%#########################################################################
%=========================================================================
\section{References} 

%% References


\bibliographystyle{elsarticle-num}
\bibliography{all}






% \end{thebibliography}

\end{document}